\documentclass[aps,prl,reprint,superscriptaddress]{revtex4-2}

\usepackage{graphicx}
\usepackage{amsmath}
\usepackage{amssymb}
\allowdisplaybreaks

\newcommand{\hs}{\hat s}
\newcommand{\hd}{\hat d}
\newcommand{\lead}[1]{\textit{#1.-}}

\begin{document}

\title{Chiral edge modes from lattice Chern-Simons theory}

\author{Gordon W. Semenoff}
\email{gordonws@phas.ubc.ca}
\affiliation{Department of Physics and Astronomy, University of British Columbia,\\
 6224 Agricultural Road, Vancouver, British Columbia, Canada V6T 1Z1}

\date{\today}

\begin{abstract}
It is shown that a level $k$ Abelian lattice Chern-Simons theory with continuum time and a spatial square lattice, which is a topological field theory on the latticized plane and torus, when quantized on a spatial cylinder reduces to chiral edge excitations, exactly localized on the two boundaries and carrying two commuting lattice versions of the level-$k$ $U(1)$ Kac-Moody algebra of opposite chirality. The algebras are represented on a Fock space in the noncompact theory and by a finite Weyl algebra in the compact one. We also observe that the theory has a natural framing of Wilson loops which exhibits the correct framing anomaly.

\end{abstract}

% 6. Generate the Title Element
\maketitle

\lead{Introduction} Chern-Simons theory \cite{deser} is a topological quantum field theory.  On a spacetime $\Sigma\times\mathbb R$, its Hilbert space is the quantization of the moduli space of flat connections on $\Sigma$, and the expectation values of its Wilson loops are isotopy invariants of framed knots \cite{witten}.  When $\Sigma$ has a boundary it acquires local, physical degrees of freedom which are chiral edge currents on $\partial\Sigma$ obeying a Kac-Moody algebra \cite{witten,elitzur,wen}.

It is a commonly held view that, if a lattice Chern-Simons theory is strictly constructed as a pure topological field theory, it cannot host gapless, chiral edge states (see \cite{xuchen,peng} and the no-go arguments \cite{berruto,kapustinspod,kapustin} on which it rests). In this Letter, we shall show that this is not the case.  There is an already well known lattice U(1) Chern-Simons theory which is a topological field theory and which we shall show hosts chiral edge states, remarkably similar to those of the continuum theory. 

Lattice formulations of Chern-Simons theory have been discussed since the 1980s \cite{luscher,frohlich,fradkin,susskind,eliezer,eliezer2,sun,chen,fiona,jacobson,jacobson2,berruto,xuchen,peng}.  The motivations are lattice simulations, applications to lattice spin and tight binding electron systems \cite{fradkin,eliezer,jain,zhk}, and the fact that the continuum theory itself is incomplete at short distances.  The self-linking of a Wilson loop must be defined by a framing \cite{witten} or some other regularization \cite{polyakov}, and a lattice theory would be expected to provide one.  We shall examine the $U(1)$ lattice Chern-Simons theory of \cite{eliezer,eliezer2}, a topological field theory whose Hilbert spaces on the latticized plane and torus are identical to those of the continuum theory.  Our central result is that, when the lattice has a boundary, it supports chiral edge states which are exactly localized on the edges and entirely determined by the bulk theory.  The derivation is a symplectic reduction of the topological field theory.  No Maxwell term or other dynamics are added.  We discuss both the compact and non-compact versions.  The flux attachment scheme of this theory has already been recognized as a framing of charged-particle trajectories, and some consequences were studied in \cite{eliezer,eliezer2}.  Here we add an observation about the framing anomaly and anyon spin.  A more complete analysis of framed knots and their invariants in this context is forthcoming \cite{long}.

For the edge states, the picture that emerges is simple.  On the latticized plane, with fixed classical sources, the lattice Chern-Simons theory has no degrees of freedom at all.  Every link variable is removed either by a constraint or by a gauge redundancy. The quantum theory has a single state whose phase records the braiding of the sources.  On a cylinder, the degrees of freedom that survive the constraints and the gauge redundancy reside on the two edges, and they carry two commuting lattice versions of the level-$k$ $U(1)$ Kac-Moody algebra of opposite chirality.  
%, with $L_1$ modes rather than infinitely many, which become the Kac-Moody algebra at long wavelength; in the compact theory only the exponentiated form of this algebra, a finite Weyl algebra, acts.

\lead{The lattice theory} The lattice theory uses continuum time (which evades the Chern-Simons doubling theorem \cite{berruto}).  Space is a two-dimensional square lattice with sites $x\in\mathbb Z^2$ and unit basis vectors $\hat 1,\hat 2$.  The gauge field $A_i(x,t)$ lives on the link $[x,x+\hat i]$ and $A_{-i}(x,t)=-A_i(x+\hat i,t)$.  $A_0(x,t)$ lives on the site $x$.   Forward and backward shift operators are $s_if(x)=f(x+\hat i)$, $\hs_i=s_i^{-1}$, and difference operators are $d_i=s_i-1$, $\hd_i=1-\hs_i$, with $\sum\limits_xf\,d_ig=-\sum\limits_x(\hd_if)g~$+ boundary terms.  The magnetic flux through the plaquette with lower-left corner $x$ is $B(x)=\epsilon_{ij}d_iA_j(x)$ with $\epsilon_{12}=1$.  The lattice Chern-Simons theory is defined by the action \cite{eliezer}
\begin{align}
S_{CS}=\frac{k}{4\pi}\int dt\sum\limits_x \Bigl\{ A_i\kappa_{ij}\dot A_j-2A_0\epsilon_{ij}d_iA_j
\Bigr\}
\label{action}
\end{align}
where coupling to a classical charge and current  would add $S_I=\int dt\sum\limits_x
(A_0\rho+A_iJ_i)$ with $\dot\rho+\hd_iJ_i=0$.
The symplectic form $\kappa_{ij}$ is a function of the shift operators with $\kappa_{ij}\delta(x,y)=-\kappa_{ji}\delta(y,x)$, where $\delta(x,y)$ is the Kronecker delta.  
Gauge invariance under the transformation $A_i\to A_i+d_i\chi$, $A_0\to A_0+\dot\chi$ requires
\begin{equation}
\hd_i~\kappa_{ij}~=~\epsilon_{ij}~d_i
\label{gauge}
\end{equation}
Equation (\ref{gauge})  is solved by
\begin{equation}
\kappa=\frac12\begin{pmatrix}\hs_2-s_2&-1+s_1+\hs_2+s_1\hs_2\\1-\hs_1-s_2-\hs_1s_2&s_1-\hs_1\end{pmatrix}
\label{kappa}
\end{equation}
which is invertible,  
\begin{equation}
\kappa^{{-1}}=\frac12\begin{pmatrix}s_1-\hs_1&1-s_1-\hs_2-s_1\hs_2\\-1+\hs_1+s_2+\hs_1s_2&\hs_2-s_2\end{pmatrix}
\label{kinv}
\end{equation}
Equation (\ref{kappa}) is the unique local solution of (\ref{gauge}) whose inverse is ultra-local, equal to $\pm\frac12$ for any two links which share a site and zero otherwise \cite{SM}.

Varying $A_0(x)$ in the action $S_{CS}+S_I$  yields the ``north-east-quarter'' Gauss' law
\begin{equation}
\frac{k}{2\pi}B(x)-\rho(x)=0
\label{Gauss law}
\end{equation}
which attaches to the charge on site $x$ the flux of the plaquette directly to the north-east of that site.
Quantization leads to the commutation relation
\begin{equation}
[A_i(x),A_j(y)]~=~\frac{2\pi i}{k}\kappa^{-1}_{ij}\delta(x,y)
\label{commutator}
\end{equation}
which is ultra-local. With this commutator,  the Gauss law constraint  (\ref{Gauss law}) generates time-independent gauge transformations  $[A_i(x),\sum\limits_y\chi(y)B(y)]=\frac{2\pi i}{k}~d_i\chi(x)$. 

The commutator (\ref{commutator}) has a simple meaning already outlined in ref.\cite{eliezer2}.  For two curves $C,C'$ made of lattice links, (\ref{commutator}) leads to
\begin{equation}
\Bigl[\sum\limits_CA,\sum\limits_{C'}A\Bigr]=\frac{2\pi i}{k}\nu(C,C')
\label{intersection}
\end{equation}
and $\nu$ is the signed intersection number of the two curves, computed with the local rules that a transverse crossing counts $\pm1$ and a meeting at a shared site without crossing counts $0$.  For two closed curves it is the homological intersection number which is zero in the plane. 
The Poisson structure in (\ref{commutator}) and (\ref{intersection}) is thus the intersection form of lattice curves, as the continuum Chern-Simons bracket is the intersection form of curves in $\Sigma$.
 
 In the continuum U(1) Chern-Simons theory, integrating out the gauge field in the presence of charged world-lines gives a phase $\frac{2\pi}{k}$ times the mutual linking numbers of distinct world-lines plus $\frac{\pi}{k}$ times the self-linking of each, and the self-linking is defined only once each world-line is thickened to a ribbon \cite{polyakov,witten}.  In the lattice theory (\ref{action}), the choice of plaquette in the north-east rule (\ref{Gauss law}) is a choice of framing.  The two edges of the ribbon are the world-lines of a charge and of its flux.  
 %The entries $\pm\tfrac12$ in (\ref{kappa}) can be seen of this choice.    
 
 A $90^\circ$ rotation of the lattice carries the flux from the north-east to the north-west plaquette, and four such rotations carry it once around the charge and change the self-linking by one unit, so the point group acts on a composite with a $2\pi$ rotation $R^4=e^{i\pi\rho^2/k}=e^{2\pi ih_\rho}$, where the spin is $h_\rho=\rho^2/2k$. The theory is invariant under a $2\pi$ rotation only up to the spin phase.  This is the lattice form of the framing anomaly \cite{witten}.  The framing, the statistics of transported charges and the amplitudes of knotted world-lines are examined in more detail in \cite{long}.

\lead{Compact and non-compact theories} The action (\ref{action}) defines two theories, compact and non-compact.  In the non-compact theory the link variables are real, the gauge group is $\mathbb R$ and $k$ is any real number.  On the torus, for example, the physical degrees of freedom are the holonomies $a_1,a_2$ around the two generators of the first homology, with (\ref{intersection}) implying $[a_1,a_2]=2\pi i/k$. If quantum states are functions of $a_2$ and $a_1=\frac{2\pi i}{k}\frac{d}{da_2}$, the Hilbert space is $L^2(\mathbb R)$, which is infinite-dimensional.  In the compact theory, the gauge fields become angle variables and $a_i\sim a_i+2\pi$.  Wavefunctions must be periodic, eigenvectors of $a_1$ are $\exp(-ina_2)$ with eigenvalues $\frac{2\pi}{k}n$. But $a_1\sim a_1+2\pi$  implies $\frac{2\pi}{k}n\sim\frac{2\pi}{k}n+2\pi$ 
which is only possible if $k$ is an integer whence 
$n$ and $n+k$ are identified, reducing the dimension of the Hilbert space to $k$, the same Hilbert space as the continuum $U(1)_k$ theory \cite{eliezer}.

On the lattice the unitary that shifts a single link $UA_1(x)U^\dagger=A_1(x)+2\pi$  is 
%$\exp\left(ik\sum_{j,y}A_j(y)\kappa_{j1}\delta(y,x)\right)$, which by (\ref{kappa}) is
\begin{equation}\begin{aligned}
U=\exp\Bigl\{ \tfrac{ik}{2}\bigl[&A_1(x+\hat2)-A_1(x-\hat2)+A_2(x)-A_2(x+\hat1)\\&-A_2(x-\hat2)-A_2(x+\hat1-\hat2)\bigr]\Bigr\}
\end{aligned}\label{shift}
\end{equation}
It involves the six other edges of the two plaquettes that contain the link, each with coefficient $\pm\tfrac k2$.  It  is periodic in the link variables only when $k$ is even.  For odd $k$ the theory contains a transparent charge-$k$ fermion and, like the continuum $U(1)_k$ theory at odd $k$, requires a spin structure \cite{long}.  Since $B\sim B+2\pi$, Gauss' law (\ref{Gauss law}) implies  $\rho\sim\rho+k$, and the gauge transformations include $\chi\to\chi+2\pi n(x)$.  The compact and noncompact theories have the same symplectic form, the same Gauss law and the same gauge transformations, so every bracket computed below holds for both, with equations read modulo $2\pi$ in the compact case.

\lead{The cylinder} We denote $(x_1,x_2)=(x,y)$.  Consider $S_{CS}$ in (\ref{action}) and let the lattice be periodic in direction $\hat 1$ with $L_1$ sites so that $x\sim x+L_1$  and have site rows $y=0,\dots,L$. 
%There are $L+1$ rows of horizontal links, $L$ rows of vertical links, $L$ rows of plaquettes.  
The cylinder theory is (\ref{action}) with the sums running over the sites and links of the cylinder. The multiplier $A_0(x,y,t)$ lives on the lower-left corner of each plaquette, that is on site rows $0,\dots,L-1$.  The top row of sites is not associated with plaquettes, $A_0(x,L,t)$ does not appear in the action and
we must impose the boundary condition $A_0(x,L,t) =0$, since otherwise the integral over a variable absent from the action would make the partition function diverge.   
No further boundary conditions are needed.  The asymmetry between the two boundary rows is that of the north-east rule (\ref{Gauss law}).  Invariance under the gauge transformation requires that the gauge function obeys $\chi(x,0,t)=0=\chi(x,L,t)$. Gauge transformations with other boundary values change the action by boundary terms.

We can use the remaining gauge symmetry to fix a gauge where $A_2$ does not depend on $y$. Gauss' law then requires $A_1(x,y+1,t)-A_1(x,y,t)=d_1A_2(x,t)$, which is solved by
$A_1(x,y,t)=A_1(x,0,t) +yd_1A_2(x,t)$. Gauss law  implies 
\begin{equation}\label{locked}
\sum\limits_x A_1(x,L,t)=\sum\limits_x A_1(x,0,t)
\end{equation}
When we plug this solution into the action, we find that it depends on the remaining undetermined variables which are conveniently written as
\begin{equation}\begin{aligned}
A_1(x,0,t)&=a(t)+d_1\chi_0(x,t)\\ A_1(x,L,t)&=a(t)+d_1\chi_L(x,t)
%\\A_2(x,t)=A_2^{(0)}(t)+\bigl[\chi_L(x,t)-\chi_0(x,t)\bigr]/L
\end{aligned}\label{edge fields}
\end{equation}
%and %$A_2(x,t)=A_2^{(0)}(t)+\bigl[\chi_L(x,t)-\chi_0(x,t)\bigr]/L$,
We find two decoupled local field theories
\begin{equation}\begin{aligned}
S=\frac{k}{4\pi}\!\int\!dt\sum\limits_x\Bigl[\chi_0\frac{s_1-\hat s_1}{2}\dot\chi_0 - \chi_L\frac{s_1-\hat s_1}{2}\dot\chi_L\Bigr]
\\
+\frac{k}{2\pi}LL_1\int dt a\dot A_2^{(0)}
\label{reduced}
\end{aligned}\end{equation}
which are copies of a discretized Floreanini-Jackiw action \cite{Floreanini:1987as}. 
It describes two chiral bosons of opposite chirality, one on each edge. Their commutators, for $L_1$ even and $|x-y|<L_1/2$, are
\begin{equation}\begin{aligned}
&\left[\chi_0(x,t),\chi_0(y,t)\right]= \frac{2\pi i}{k}\mathrm{sgn}(x-y) \frac{1-(-1)^{x-y} }{2}
\\&\left[\chi_L(x,t),\chi_L(y,t)\right] = - \frac{2\pi i}{k}\mathrm{sgn}(x-y) \frac{1-(-1)^{x-y} }{2}
\\&\left[\chi_0(x,t),\chi_L(x',t)\right]=0
\\&\left[a(t),A_2^{(0)}(t)\right]=-\frac{2\pi i}{LL_1k}
\end{aligned}\label{chiral boson ccr}
\end{equation}
The last commutator  in (\ref{chiral boson ccr}) is equivalent to  
\begin{equation}\label{zero modes}
\left[\sum_xA_1(x,0,t),\sum_yA_2(x',y,t)\right]=-\frac{2\pi i}{k}
\end{equation}
 The Wilson line which wraps the cylinder is conjugate to the Wilson line across it, consistent with (\ref{intersection}). 
The cross-line operator is the product of 
 top-edge and bottom-edge vertex operator with the zero-mode shift  $\exp(i\sum\limits_yA_2(x,y))=e^{iLA_2^{(0)}}e^{i\chi_L(x)}e^{-i\chi_0(x)}$. It transfers one unit of charge from one edge to the other through the bulk, and it is the only thing that couples the edges, exactly as in the continuum theory on an annulus.

In the non-compact theory the edge Hilbert space is the infinite dimensional Fock space of the two lattice chiral bosons, and the zero modes (\ref{zero modes}) are represented by $L^2(\mathbb R)$.  In the compact theory the observables are the exponentials $e^{inA_1(x,0)}$, and (\ref{chiral boson ccr}) gives the clock algebra
\begin{equation}
e^{iA_1(x,0)}e^{iA_1(y,0)}=e^{\frac{4\pi i}{k}(-1)^{x-y}\mathrm{sgn}(x-y)}e^{iA_1(y,0)}e^{iA_1(x,0)}
\label{vertex}
\end{equation}
Each edge is a compact chiral boson at level $k$.  On a finite lattice the operators (\ref{vertex}) generate a finite Weyl algebra, so the edge Hilbert space is finite-dimensional, with the dimension growing exponentially with the length of the edge, the chiral boson of (\ref{reduced}) being its long-wavelength form.  The Hilbert space of the cylinder has the sector structure of continuum Chern-Simons theory on an annulus. By (\ref{zero modes}) the holonomy and the cross line form a $\mathbb Z_k$ clock and shift pair, so it is a diagonal sum over $k$ sectors, $\bigoplus_{j=0}^{k-1}\mathcal H_j\otimes\bar{\mathcal H}_j$, with the two edges in the same sector.  The factors are not the Kac-Moody modules of the continuum edge, which are infinite-dimensional.  $\mathcal H_j$ is a finite-dimensional representation of the Weyl algebra (\ref{vertex}), the exponentiated form of the current algebra, which is represented as a Lie algebra only in the non-compact theory.  The operators (\ref{vertex}) are the lattice edge vertex operators, the charge-one anyons of the edge, and the edge charge is measured by the holonomy $e^{i\sum\limits_xA_1(x,0)}=e^{-2\pi iQ/k}$, modulo $k$, in agreement with the bulk charge modulo $k$.

\lead{Edge charge and anomaly} We have established that the degrees of freedom on the cylinder are $a,A_2^{(0)},\chi_0,\chi_L$ in (\ref{reduced}) and (\ref{chiral boson ccr}).  We will now show that these modes are electrically charged, with the charge densities $q(x,0)$ and $q(x,L)$ given in (\ref{q}) below, that the densities obey the level-$k$ current algebra and that this algebra is protected by an anomaly.
In a Chern-Simons theory charge is defined by Gauss' law. When summed over the plaquettes inside a loop, (\ref{Gauss law}) says that the charge enclosed is $\frac{k}{2\pi}$ times the counterclockwise holonomy.  Every loop in the bulk of the cylinder has zero holonomy.  The bulk is neutral.  
 
 Consider the loop made of a horizontal link $A_1(x,y)$ in the bulk, the two columns of vertical links joining its ends to the bottom $y=0$  boundary and continued one step beyond it, and a horizontal link below the boundary, with the gauge field set to zero on the links outside the lattice.  By Gauss' law its holonomy is independent of $y$, so it can be evaluated at $y=0$, where it equals $-A_1(x,0)$.  The enclosed charge, $-\frac{k}{2\pi}A_1(x,0)$, is therefore concentrated on the boundary link.  This is the charge that Gauss' law requires at the boundary, and the Noether bracket below confirms it.  By the same argument the top boundary links carry  charge $\frac{k}{2\pi}A_1(x,L)$, and with (\ref{locked}) the total charges of the two edges add to zero.  Per site, the charge densities are
\begin{equation}\begin{aligned}
q(x)&\equiv-\frac{k}{2\pi}\cdot\frac{A_1(x,0)+A_1(x-1,0)}{2}\\&=-\frac{k}{2\pi}a-\frac{k}{4\pi}\bigl[\chi_0(x+1)-\chi_0(x-1)\bigr]
\end{aligned}\label{q}
\end{equation}
with the opposite sign on row $L$. The edge charge density is the symmetric lattice difference of the chiral boson (analogous to $\frac{k}{2\pi}\partial_x\phi$ in the continuum).  From (\ref{chiral boson ccr}), $[q(x),\chi_0(x')]=-i\delta_{xx'}$  and $\sum_x\xi_0q$ generates the shift $\chi_0\to\chi_0+\xi_0$, which is a gauge transformation with boundary value $\xi_0$. A constant $\xi_0$ is the global $U(1)$ transformation of the edge and $q$ its Noether density.  Since $q$ is itself a difference of $\chi_0$'s, the generators do not commute, on row 0,
\begin{equation}
[q(x),q(x')]=-\frac{ik}{2\pi}\cdot\frac12\bigl(\delta_{x',x+1}-\delta_{x',x-1}\bigr)
\label{km}
\end{equation}
with the opposite sign on row $L$, and those on rows 0 and $L$ commute with each other. This is a lattice level-$k$ $U(1)$ Kac-Moody algebra, with $\delta'(x-x')$ replaced by the symmetric lattice difference. For a $U(1)$ current algebra the level is the normalization of the central term relative to the unit of charge, so it is any real number in the non-compact theory and an integer in the compact one, where (\ref{vertex}) requires it.  In momentum space $[q(p),q(-p)]=\frac{k}{2\pi}\sin p>0$ on $0<p<\pi$. Each edge is therefore chiral over the whole Brillouin zone whatever Hamiltonian is chosen. % The zero at $p=\pi$ is the staggered mode of $\chi_0$, a redundancy that $q$ annihilates.

Secondly, the gapless chiral edge states are protected by a 't Hooft anomaly.  The edge states that we have found have no dynamics until a Hamiltonian is added in the form of local boundary terms.  Protection is a statement about the algebra (\ref{km}).  Nothing done at the boundary can change its level or gap it.  The boundary data are the gauge-invariant links $A_1(x,0)=a+d_1\chi_0$, since $A_0$ vanishes on the top row and multiplies a constraint on the bottom one and $A_2$ is not gauge invariant, and every local boundary term is a local functional of them.  A potential does not change (\ref{km}).  A source $A_0f(A_1)$ only adds $-f$ to the edge charge.  A term $A_1F\dot A_1$ changes the symplectic form but, being local in $A_1$, it is local in $\chi_0$ with two extra differences and alters (\ref{km}) only at order $p^3$.  The level is the coefficient of (\ref{reduced}), which is local in the gauge parameter $\chi_0$ but non-local in the gauge field.  It is the change of the bulk action under a gauge transformation with boundary value $\chi_0$, and that is what it means for it to be an anomaly.  The boundary cannot cancel the variation of the bulk action.  Nor can it gap the edge.  For any local quadratic Hamiltonian $\sum_xqf(s_1,\hs_1)q$ the frequencies are $\omega(p)\propto f(p)\sin p$, which vanish at $p=0$ whatever $f$ is.  For any local Hamiltonian that conserves the charge, the central term gives $q(\pm p)|0\rangle$ a weight of order $p$ above the ground state, while $\langle[q(-p),[H,q(p)]]\rangle$, which bounds their energy, is of order $p^2$, so there is no gap.  In this sense the topological theory has protected gapless chiral edge states.  In the compact theory $q$ is the sum of two link variables and shifts by $k/2$ under $A_1\to A_1+2\pi$ on one link, so $q$ itself is not an operator.  Only $e^{2\pi i(2q)/k}=e^{i[A_1(x,0)+A_1(x-1,0)]}$ is.  The vertex operators (\ref{vertex}) fuse additively, $e^{inA_1}e^{imA_1}=e^{i(n+m)A_1}$, and form a projective representation of the abelian group of edge charges whose cocycle is the exponential of (\ref{km}).  In the compact theory it is this Weyl form of the current algebra that exists, rather than a Lie algebra of operators, as for any compact lattice boson \cite{jacobson}.

\lead{Edge dynamics} A current appears once the edge is given a velocity.  The natural edge energy is the discretization of the continuum $\frac{kv}{4\pi}\int A_x^2$, local in the link variable,
\begin{equation}
H=\frac{kv}{4\pi}\sum\limits_xA_1(x,0)^2
\label{wave}
\end{equation}
and with it the edge charge obeys an exact lattice continuity equation,
\begin{equation}
\dot q(x)+\hd_1j(x)=0,\qquad j(x)=\frac{kv}{2\pi}\,A_1(x,0)
\label{current}
\end{equation}
density on sites and current on links, the current being the boundary link variable itself and $j\simeq-vq$ at long wavelength.  The dispersion is the ratio of the two lattice differences that the bulk has supplied.  The symmetric difference in the symplectic form gives $i\sin p$, because a symplectic form must be antisymmetric, and the forward difference in the energy gives $|1-e^{ip}|^2=4\sin^2(p/2)$, because the link variable is a forward difference.  Hence
\begin{equation}
\omega(p)=v\,\frac{4\sin^2(p/2)}{\sin p}=2v\tan\frac p2
\label{dispersion}
\end{equation}
a single branch, monotonic from $\omega\simeq vp$ to infinity at the zone boundary, with no counter-propagating mode. The one-dimensional doubling expected of a local lattice theory \cite{nielsen} does not occur.  A Hamiltonian local in $q$ would give $\omega=h(p)\sin p$, which returns to zero at $p=\pi$ and carries no net chirality, as it must for any variable with a local commutator.  The non-locality that \cite{nielsen} requires is in the commutator of the link variable, which the bulk supplies through the constraints, while the energy is a sum of single-link terms.  This pair, a field with the forward difference and a charge with the symmetric difference, is the discretization proposed by Stacey to evade fermion doubling \cite{stacey} and more recently revived as the ``tangent fermion'' \cite{pacholski,tangent,zakharov}. There it is chosen because it works, here it is what a topological bulk produces at its boundary.   In the compact theory the quadratic Hamiltonian (\ref{wave}) is not periodic. The edge Hamiltonian is a cosine in the link variables. Chirality is then carried by the symplectic form and by (\ref{vertex}), as for the continuum chiral Luttinger liquid, rather than by a free dispersion.

\lead{Discussion} The objection that lattice Chern-Simons theory cannot be chiral rests on three arguments,  the doubling of parity-odd kernels on spacetime lattices \cite{berruto}, the periodicity argument that a local one-dimensional lattice algebra cannot be chiral \cite{nielsen}, and the absence of Hall conductance in commuting-projector Hamiltonians \cite{kapustinspod,kapustin}.  These do not apply to the present theory.  Its symplectic form (\ref{kappa}) is nondegenerate. There are no spurious zero modes to gap.  Its edge is the leftover of a two-dimensional symplectic reduction, and the non-locality the doubling theorem demands is in the commutator of the boundary link variables.  It is first order in time, with a Hilbert space that is not a tensor product over sites, so the commuting-projector theorems do not apply, as in \cite{demarco}.  The chirality is a property of the symplectic form inherited from the sign of $k$.

The comparison with lattice Chern-Simons-Maxwell theory \cite{xuchen,peng} is instructive.  There the Maxwell term is held to be ``necessary for the lattice realization to have the correct local dynamics'' and the edge modes are ``not entirely localized at the boundary'' and ``always merge into the bulk modes''.  Here there is no Maxwell term, the bulk has no local dynamics at all, and the edge modes are exactly localized on the boundary rows and independent of the width of the cylinder.  The characterization in \cite{xuchen} of the present theory as lacking a locally factorized Hilbert space with a local Gauss law is correct, and it is the mechanism.  A theory with a locally factorized Hilbert space structure is a commuting-projector model and cannot be chiral.   The coarse-graining argument of \cite{xuchen}, that a purely topological lattice theory could be blocked down to a few cells, which could not host a chiral edge, fails at the same point. Blocking presupposes a Hilbert space that factorizes over cells, and the cylinder with a single row of plaquettes already carries the two edges, with the same algebra, level and dispersion as any wider one.  The characterization of its symplectic form as fine-tuned is unfair. It is determined by the intersection form and the framing, and the edge does not depend on the remaining freedom.  Any other local solution of (\ref{gauge}) adds to the action the term $-\frac{k}{4\pi}\sum\limits_xB\,T\dot B$ with $T$ a local odd kernel \cite{SM}, which vanishes when Gauss' law holds, so (\ref{reduced}), (\ref{q}) and (\ref{km}) are unchanged.

\section{Supplement: Uniqueness of the symplectic form}

Equation numbers of the form (2), (3), (7) refer to the Letter.  Let $\kappa'$ be any local antisymmetric solution of the gauge condition (2) and $\Delta=\kappa'-\kappa$, with $\kappa$ the symplectic form (3).  Then $\hd_i\Delta_{ij}=0$, and by antisymmetry $\Delta_{ij}d_j=0$, whose general local solution is
\begin{equation}
\Delta_{ij}=(\epsilon\hd)_i\,T\,(\epsilon d)_j
\label{Delta}
\end{equation}
with $T$ a local scalar, odd under $s_i\leftrightarrow\hs_i$ \cite{long}.  Since (\ref{Delta}) has rank one, $\det(\kappa+\Delta)=\det\kappa\,[1+T(\epsilon d)^T\kappa^{-1}(\epsilon\hd)]=1$, because $\kappa^{-1}\epsilon\hd=d$ by (2) and $(\epsilon d)^Td=\epsilon_{ij}d_jd_i=0$.  Every $\kappa'$ is unimodular, with the local inverse
\begin{equation}
\kappa'^{-1}_{ij}=\kappa^{-1}_{ij}+T\,d_i\hd_j
\label{Delta inverse}
\end{equation}
An odd $T\neq0$ contains a shift $s^m$, $m\neq0$, and then $T d_1\hd_1=T(s_1-2+\hs_1)$ contains the shift $s^{m+\hat1}$, which couples $A_1(x)$ to a link $A_1(x+m+\hat1)$ that does not share a site with it unless $m=-\hat1$.  In that case $T=t(\hs_1-s_1)$ and $T d_1\hd_1=t[2(s_1-\hs_1)+\hs_1^2-s_1^2]$ couples links two apart.  Hence (3) is the only solution whose inverse couples only links that share a site.

In the action, (\ref{Delta}) is the gauge-invariant term $-\frac{k}{4\pi}\sum\limits_xB\,T\dot B$, which does not involve $A_0$ and so leaves Gauss' law unchanged.  When Gauss' law holds it is the contact term $-\frac{\pi}{k}\sum\limits_x\rho\,T\dot\rho$ between charges within the range of $T$ of each other.  For chains $C,C'$ with link occupation numbers $m_i,m'_i=0,\pm1$, so that $\sum_CA=\sum_xm_iA_i$, the change in the bracket (7) is
\begin{equation}
-\frac{2\pi i}{k}\sum\limits_x(\hd_im_i)\,T\,(\hd_jm'_j)
\label{change}
\end{equation}
Since $\hd_im_i$ vanishes except at the endpoints of a chain, where it is $\pm1$, (\ref{change}) is non-zero only when an endpoint of $C$ lies within the range of $T$ of an endpoint of $C'$.  For closed curves it vanishes, so the braiding phases and the torus algebra, which are computed from closed curves, are the same for every $T$.

\begin{acknowledgments}
This work is supported in part by the Natural Sciences and Engineering Research Council of Canada.
\end{acknowledgments}

\end{document}